# Personalization, Cognition, and Gamification-based Programming Language Learning: A State-of-the-Art Systematic Literature Review

[1]Kashif Ishaq, [1]Atif Alvi
[1]School of Systems and Technology, University of Management and Technology, Lahore

**Corresponding Author:**
kashif.ishaq@umt.edu.pk

ABSTRACT

Programming courses in computing science are important because they are often the first introduction to computer programming for many students. Many university students are overwhelmed with the information they must learn for an introductory course. The current teacher-lecturer model of learning commonly employed in university lecture halls often results in a lack of motivation and participation in learning. Personalized gamification is a pedagogical approach that combines gamification and personalized learning to motivate and engage students while addressing individual differences in learning. This approach integrates gamification and personalized learning strategies to inspire and involve students while addressing their unique learning needs and differences. A comprehensive literature search was conducted by including 81 studies that were analyzed based on their research design, intervention, outcome measures, and quality assessment. The findings suggest that personalized gamification can enhance student cognition in programming courses by improving motivation, engagement, and learning outcomes. However, the effectiveness of personalized gamification varies depending on various factors, such as the type of gamification elements used, the degree of personalization, and the characteristics of the learners. This paper provides insights into designing and implementing effective personalized gamification interventions in programming courses. The findings could inform educational practitioners and researchers in programming education about the potential benefits of personalized gamification and its implications for educational practice.

## 1. Introduction

Computer science-related fields are expected to experience the highest job growth in STEM, according to many institutions (Venter, 2020). However, students often perceive programming courses as challenging and overwhelming, negatively affecting their motivation, engagement, and learning outcomes. Researchers have found that computer programming taught in the traditional teacher-lecturer learning format needs more interaction and often results in a considerable decrease in attention (Arif, Rosyid and Pujianto, 2019). The knowledge presented in the conventional classroom is often abstract and conveyed through text or speech, which may limit students' understanding based on their prior programming experience. In recent years, two key approaches have emerged as promising strategies for enhancing programming language education: personalization and gamification.

Gamification involves using game-like elements to make learning more engaging and motivating. Educators can make programming language education fun and enjoyable by incorporating gamification techniques such as points, badges, and leaderboards (Imran, 2019). Personalized gamification takes this further by customizing the gamification elements to the individual learner's preferences, learning styles, and skill levels. Studies utilizing personalization techniques tend to group students into strict categories based on their characteristics. This method of personalization needs to be revised as it sets a constant configuration at the beginning of the study, which only allows for mutability as students progress through the course. A clear advantage of progressive personalization is that students with a good knowledge of the course material will have a personalized experience tailored to their needs, which will inherently differ from the personalized experience of programming beginners. However, these techniques' effectiveness depends on their alignment with the cognitive processes involved in learning and problem-solving. Therefore, a thorough understanding of how gamification and personalization can support cognitive processes in programming language education is essential.

This systematic literature review offers a comprehensive overview of the current state-of-the-art programming language education, specifically in personalization, gamification, and cognition. The review critically examines existing literature on the techniques, frameworks, and their effectiveness in supporting cognitive processes in programming

ORCID(s):

language education and discusses the challenges and opportunities associated with their implementation. This review also identifies gaps in the literature that require further investigation. The findings of this review have important implications for educational practice, as they inform the design and implementation of effective personalized gamification interventions in programming courses. Particularly, this review highlights the role of cognition in programming language education. Cognition refers to the mental processes involved in learning, problem-solving, and decision-making, such as attention, memory, perception, reasoning, and problem-solving. Understanding how these cognitive processes work can help educators design effective programming language curricula and interventions that promote learning and problem-solving.

The remainder of this paper is organized as follows. Section 2 provides a background on the relevant literature, including personalization and gamification in education, cognitive skills, and programming language education. Section 3 outlines our methodology for conducting the systematic literature review. Section 4 presents the results of our review, including an analysis of the key themes and trends in the literature. Section 5 discusses the implications of our findings and identifies opportunities for future research. Finally, Section 6 concludes the paper and summarizes our key contributions.

## 2. Background

Programming language education focuses on teaching students how to code and develop software and has become increasingly important as technology advances. While gamification has been widely adopted in many educational settings, the research on personalized gamification for programming courses is still relatively limited compared to non-personalized gamification. The main methodologies include creating an online learning system that personalizes which gamification elements are available by placing students into predetermined groups (Santos, Oliveira, Hamari, Rodrigues, Toda, Palomino and Isotani, 2021). The biggest gap in this area of research is a need for more fluidity or flexibility with personalization; students are statically placed in predetermined groups and need the opportunity to change based on generated information (Rodrigues, Palomino, Toda, Klock, Oliveira, Avila-Santos, Gasparini and Isotani, 2021). The current state of knowledge is that non-personalized gamification does not certainly improve cognitive abilities Sanmugam, Abdullah and Zaid (2014). Additionally, non-personalized gamification is more effective when negative emotions are elicited. The research techniques generally encompass a stronger psychological than technological standpoint (Mullins and Sabherwal, 2018, 2020).

### 2.1. Gamification

Gamification has become a popular approach to address the issue of low motivation and engagement in e-learning platforms. Researchers have conducted various studies to explore the impact of gamification on students' motivation and engagement in higher education programming courses. (Venter, 2020) conducted a systematic literature review of gamification in higher education programming courses. Points, achievements, levels, leaderboards, and badges were the most commonly used gamification elements in the reviewed studies. The studies reviewed by Venter showed that gamification positively affected engagement, motivation, and learning outcomes. Gamification in a programming language using game design elements can increase learner engagement, motivation, and retention, resulting in improved performance as learners spend more time studying and earning badges (Imran, 2022).

Similarly, (Kiraly and Balla, 2020) developed a learning management system containing online programming language courses and added points, incentives, badges, immediate feedback, and a leaderboard to gamify the courses. Their results showed that the students who completed a Java course with gamification were better at solving coding tasks.

Furthermore, (Katan and Anstead, 2020) developed a gamification platform called "Sleuth" that teaches introductory programming. They found that students who used Sleuth received a very high median grade (90.67%), while students who used a module-based testing environment received quite a relatively low grade (66.94%). Their research filled a gap in the literature by creating a gamification platform that resembles a video game. (Pankiewicz, 2020) found that gamification elements such as points, badges, and leaderboards positively impacted students' motivation in the learning process.

Additionally, (Queirós, 2019) presented a framework called "PROud" that applies gamification features based on the usage data of programming exercises, such as fostering competition between students based on the correctness of code solutions submitted. Moreover, (Hassan, Habiba, Majeed and Shoaib, 2021) investigated why students lack motivation in e-learning platforms and concluded that it stems from their learning experience.

The studies suggest that gamification elements in e-learning platforms can significantly improve students' motivation and engagement in programming courses.

## 2.2. Personalization

In recent years, researchers have emphasized the significance of personalization in gamification research. It has been suggested that considering users' characteristics can enhance the potential benefits of gamification (Rajanen and Rajanen, 2017; Ghaban and Hendley, 2019). Personalization can increase students' motivation and engagement by providing a tailored learning experience that caters to their needs, preferences, and performance. Adaptive learning, personalized feedback, and customization of game elements based on students' profiles are some of the gamification techniques that can be used to achieve personalization.

(Challco, Moreira, Bittencourt, Mizoguchi and Isotani, 2015) conducted a study on using ontologies to personalize gamification in collaborative learning environments. They proposed that collaborative gamification techniques could address the issue of decreased motivation. Similarly, (González, Toledo and Muñoz, 2016) explored enhancing student engagement in learning systems through the personalization of gamification. They developed an intelligent tutorial system that incorporated adaptation and personalization of gamified elements.

(Roosta, Taghiyareh and Mosharraf, 2016) focused on personalizing gamified elements in an online learning environment based on learners' motivation. Their study proposed characterizing various game elements and students' motivation types to create a personalized learning management system. The personalization system adapted the gamified elements displayed to students based on their motivation category. (Knutas, Roy, Hynninen, Granato, Kasurinen and Ikonen, 2017) designed a profile-based algorithm for personalization in online collaborative learning environments based on intrinsic skill atoms and gamification-based user-type heuristics. They also developed personalized gamification software using this profile-based algorithm.

Several studies have explored the benefits of personalized gamification on student engagement, motivation, and cognition. (Knutas, van Roy, Hynninen, Granato, Kasurinen and Ikonen, 2019) developed a machine learning-based personalized content system. (Rodrigues, Toda, Oliveira, Palomino, Vassileva and Isotani, 2022) investigated the relevance of personalization characteristics and collected user feedback on game elements such as points and rewards. (Bennani, Maalel and Ghezala, 2020) created an adaptive gamification ontology called "AGE-Learn" and found that personalized gamification improved online student engagement, motivation, and cognition. (Santos et al., 2021) grouped users into six categories and studied the association between user types and their feedback on different gamification elements.

Personalization is particularly beneficial because students have unique learning styles, personalities, values, and motivating factors. Overall, personalization is a key aspect of gamification research, and several studies have explored how it can be used to enhance the potential benefits of gamification for learners.

## 2.3. Cognition and Gamification

Research has shown that gamification can positively affect cognitive processes, such as attention, memory, and learning (Hamari, Koivisto and Sarsa, 2014). Gamification can enhance cognitive engagement by encouraging active information processing, focused attention, and decision-making based on feedback provided by the game. Several studies have found that gamification can improve cognitive engagement and learning outcomes.

For instance, (Rojas-López, Rincón-Flores, Mena, García-Peñalvo and Ramírez-Montoya, 2019) explored gamification's impact on engagement in higher education programming courses. Their study emphasized gamification's emotional and social aspects, stating that recognizing students for their accomplishments through awards, trophies, or achievements can provide emotional motivation, and encouraging students to work together to complete a task can provide social motivation. The results indicated that gamification significantly improved student engagement.

Furthermore, (Mullins and Sabherwal, 2020) approached gamification from a cognitive-emotional perspective. They highlighted the importance of considering both positive and negative emotions in gamification. They suggested that emotions and cognitions can interact further to enhance the positive outcomes of a gamified system.

Recently, there has been growing interest in exploring the potential of combining these three areas to improve the effectiveness of programming language education. Educators hope to increase student engagement and motivation by personalizing instruction and incorporating gamification elements, promoting more effective learning outcomes. Additionally, understanding the role of cognition in programming language education is crucial for designing effective curricula and interventions. Educators can design interventions that promote more effective learning and problem-solving in programming language education by understanding how these processes work and how they can be supported.

Previous research has explored various aspects of personalization, gamification, and programming language education and has highlighted the potential benefits of each. However, no systematic literature review examines the state-of-the-art in personalization, gamification, cognition, and programming language education and how these areas intersect. Our current paper aims to fill this gap by providing a thorough overview of the existing literature and identifying areas for further research.

## 3. Methodology

This section outlines the systematic literature review process to identify and analyze relevant studies on personalized gamification, cognition, and programming language education. It describes selecting appropriate search terms, databases, and inclusion and exclusion criteria. Furthermore, it outlines the screening and selection procedure, as well as the techniques employed for data extraction and quality assessment to ensure the dependability and accuracy of the results.

The following subsection outlines the research questions.

### 3.1. Research questions

- RQ1 aimed to establish an article library on personalized gamified programming education and make the dataset available to other researchers. Additionally, it sought to explore how personalized gamified programming education can enhance students' cognitive abilities. The findings of RQ1 identified trends, publication channels, and geographical areas covered in the articles.
- RQ2: What criteria can be used to assess the quality of papers selected for review in the context of personalized gamified programming education, and how can these criteria be applied to the articles identified in RQ2 to ensure that only high-quality research is included in the dataset?
- RQ3: What are the trends and best practices for implementing personalized gamification frameworks in programming education, and how do these frameworks differ in design and customization?
- RQ4: How can personalized gamification frameworks in programming education be mapped to the different cognitive levels of Bloom's taxonomy?
- RQ5. What tools and software applications are developed based on personalized gamification frameworks in programming education, and how are these tools tailored to specific programming languages and concepts?
- RQ6: What are the common processes, tools, and instruments utilized to evaluate applications based on personalized gamified programming education? What evaluation measures are employed to assess applications from various viewpoints, such as teaching, learning, and technical perspectives?

### 3.2. Search Strategy

The search strategy for this study was designed to identify relevant articles based on the research questions. The specific details of the search strategy are presented in the following subsections.

#### *3.2.1. Automated Search in Web of Science (WoS Core Collection)*

A systematic investigation was carried out to filter irrelevant research and obtain adequate information. The Web of Science Core Library is a curated database of over 21,100 peer-reviewed journals, including top-tier academic journals worldwide (including Open Access journals), covering over 250 disciplines (University). It is widely regarded as a tool that helps users efficiently gather, analyze, and share information from various databases (Clarivate).To conduct the Systematic Literature Review (SLR) in an organized and efficient manner, the researcher utilized this platform to retrieve research articles by combining AND and 'OR' Boolean operators with keywords to create a search string. Figure 1 provides an overview of the search results obtained from the Web of Science. Table 1 presents the ultimate search string, which utilized 'AND' and 'OR' Boolean operators with keywords to query the WoS Core Collection. The search was limited to titles only, and a filter based on indices and time span was applied to narrow down the search query for the study.

**Table 5**
Search Strategy for Digital Library

| Digital Library | Search Query | Applied Filter |
|---|---|---|
| (WoS Core Collection) SCI-Expanded SCIE ESCI AHCI | gamification (Title) OR gamified (Title) OR game (Title) OR game-based (Title) OR game-based (Title) OR serious game (Title) AND programming (Title) OR programming (Title) OR programming course (Title) OR programming subject (Title) AND Cognition (Title) OR comprehension (Title) OR perception (Title) OR understanding (Title) OR learning (Title) AND personalization (Title) OR realization (Title) OR actualization (Title) | 2015-2022 |

### *3.2.2. Inclusion Criteria*

The paper included in the review must be in the domain of personalized, cognition, and game-based computer programming learning that must target the research questions. The paper published in journals or conferences from 2015 through 2022 is included in the review.

### *3.2.3. Exclusion Criteria*

Papers excluded from the study that was not written in English were not accessible, and also that do not discuss or focus on personalized, cognition, and game-based computer programming in educational institutes. A detailed flowchart of inclusion/exclusion criteria is presented in Figure 1.

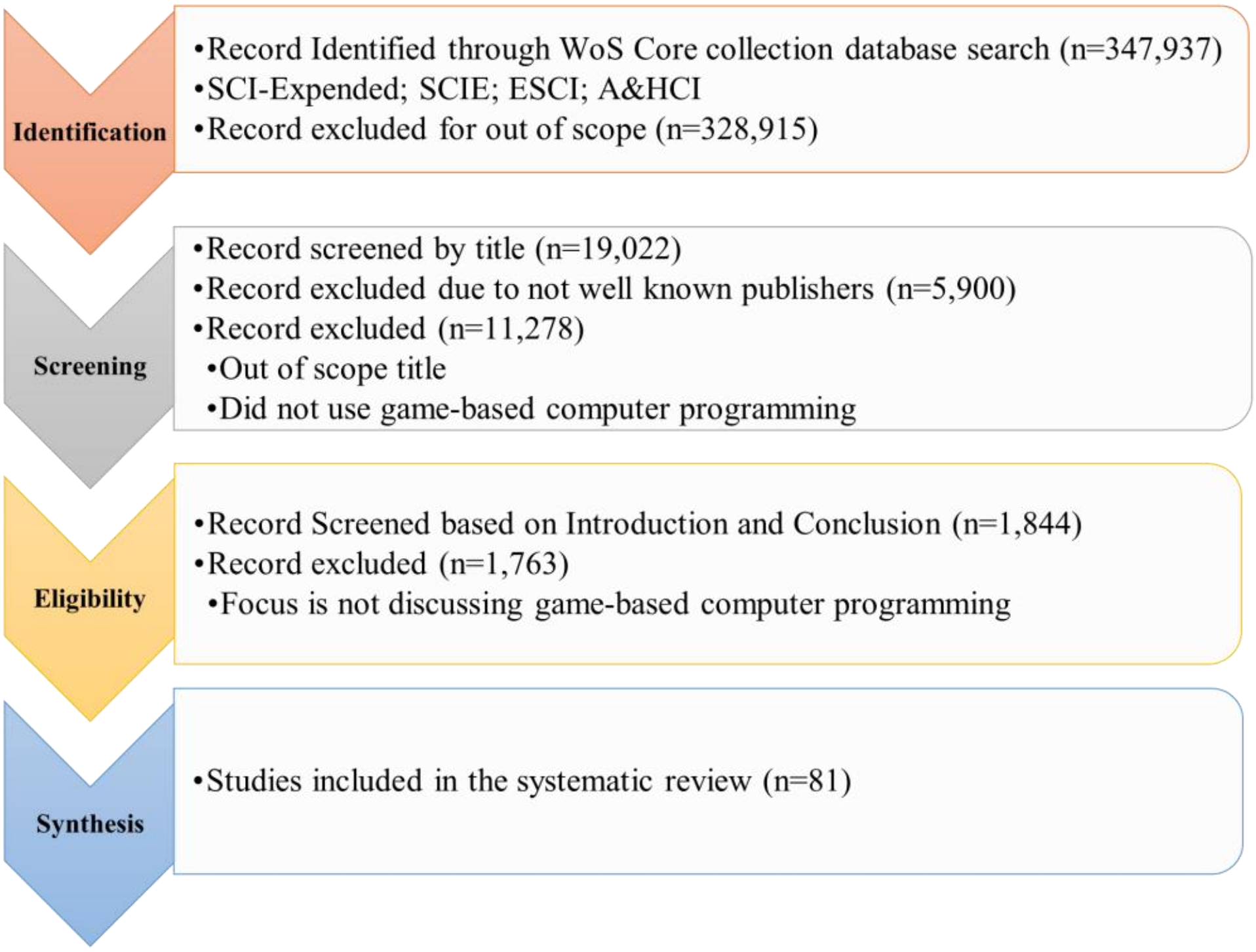

**Figure 1:** Flow chart of Systematic Review Process

## 3.3. Screening Process

The screening process consisted of two stages: title and abstract screening and full-text screening. Two reviewers independently screened the titles and abstracts of all identified articles against the inclusion criteria. After title and abstract screening, two independent reviewers retrieved and reviewed full-text articles against the inclusion criteria. Any discrepancies were resolved through discussion between the two reviewers. The screening process followed the PRISMA guidelines (Ishaq, Zin, Rosdi, Jehanghir, Ishaq and Abid, 2021) and is presented in a flow diagram in Figure 1.

**Table 6**
Possible rating for recognized and stable publication source

| Sr. No. | Publication Source | 4 | 3 | 2 | 1 | 0 |
|---|---|---|---|---|---|---|
| 1 | Journals | Q1 | Q2 | Q3 | Q4 | No JCR Ranking |
| 2 | Conferences | Core A * | Core A | Core B | Core C | Not in Core Ranking |

## 3.4. Data Extraction

Two independent reviewers conducted the data extraction process following the PRISMA guidelines. The reviewers utilized a pre-designed data extraction Excel sheet to collect relevant information from the selected articles. The data extraction form included the following information:

- Study characteristics: authors, year of publication, title, journal/conference, country, research design, sample size, and study duration.
- Gamification and personalization features: gamification elements used, personalization techniques applied, and their effects on learning outcomes.
- Cognitive aspects: the impact of gamified and personalized programming education on cognitive skills, such as problem-solving, critical thinking, creativity, and motivation.
- Programming languages: the programming languages and concepts used in the studies.
- Evaluation methods: the evaluation methods used to measure the effectiveness of gamified and personalized programming education.

### *3.4.1. Selection based on quality assessment*

The collection of appropriate studies based on quality assessment (QA) is considered the key step for carrying out any review. As the key studies differ in nature, the critical assessment tool (Fernandez, Insfran and Abrahão, 2011) and (Ouhbi, Idri, Fernández-Alemán and Toval, 2015) used to conduct QA is also supplemented in our analysis by quantitative, qualitative, and mixed approaches. To enhance the rigour of our study, we developed a QA (quality assurance) questionnaire to assess the accuracy of the selected records. The authors conducted the QA for our research using the following parameters for each study:

a) If the analysis leads to personalized, cognition, and game-based computer programming language learning, the result is indeed (1), otherwise (0).
b) As simple answers are given for personalized, cognition, and game-based computer programming language learning, the analysis will provide the following scores: 'Yes (2),' 'Limited (1),' and 'No (0).'
c) If the studies provide empirical results, then award (1) or else score (0).
d) Studies have been analyzed in regard to graded rankings of journals and conferences in computer science (Ishaq et al., 2021). Table 2 indicates potential findings for publications from known and reliable sources.

After combining the number of above-mentioned parameters, a final score was determined for each study: (value between 0 and 8). Papers with 4 or more ratings were included in the final results.

### *3.4.2. Selection based on snowballing*

After conducting a standard appraisal, we utilized backward snowballing through the reference lists of any completed analyses to identify additional relevant papers (Mehmood, Abid, Farooq and Nawaz, 2020). Only those candidate articles that satisfied the inclusion/exclusion criteria were considered. The inclusion/exclusion of an article was determined after reviewing its introduction and other relevant sections.

**Table 3**
Selection Phases and Results

| Phase | Selection | Selection Criteria | Indexes: SCI-EXPANDED, SSCI, A&HCI, ESCI |
|---|---|---|---|
| 1 | Search | Keywords (Figure. .........) | 347,937 |
| 2 | Filtering | Title | 19,022 |
| 3 | Filtering | Abstract | 11,278 |
| 4 | Filtering | Introduction and Conclusion | 1,844 |
| **5** | **Inspection** | **Full Article** | **81** |

## 4. Data analysis

In this section, the overview of finalized studies is provided.

### 4.1. Overview of intermediate selection process outcome

Game-based programming language learning is a very active topic, and the analysis approach of the researchers is to find out suitable research systematically and empirically from the Web of Science core collection. The further step after finding the relevant research is to compile the record to form the foundation for analysis. More than 300,000 articles were found in the Web of Science core collection by providing the keywords from 2015 to 2022. Inclusion and exclusion criteria were defined for filtering the record based on titles, the abstract, article written in English, accessibility of the document, and considering well know publishers. Moreover, papers focused on personalized, cognition, and game-based computer programming language in educational institutes were included in this research, whereas the non-availability of any area in the article was excluded.

### 4.2. Overview of selected studies

Table 3 presents significant results of primary search, filtering and review processes that include Web of Science indices. At the filtering/inspection stage, this amount was decreased to 81 articles by the automatic search.

## 5. Assessment and discussion of research questions

In this section, we analyzed 81 primary studies based on our research questions. The following section presents the findings of the systematic literature review on personalized gamified programming education:

**RQ1 aimed to establish an article library on personalized gamified programming education and make the dataset available to other researchers. Additionally, it sought to explore how personalized gamified programming education can enhance students' cognitive abilities. The findings of RQ1 identified trends, publication channels, and geographical areas covered in the articles.**

Table 4 and Figure 2 present the geographical distribution of selected studies. It has been observed that most of the studies were from Europe, i.e. 36, whereas the American countries published 24 studies. Asian countries published 18 studies, while only three have been published by the ocean and the African continent.

Table 4: Publication Source

| Sr. No. | Publication Sources | Channel | No. of Articles |
|---|---|---|---|
| 1 | Education and Information Technologies | Journal | 6 |
| 2 | Interactive Learning Environments | Journal | 3 |
| 3 | Computers Education | Journal | 3 |

Continued on next page

Table 4: Publication Source (Continued)

| | | | |
|---|---|---|---|
| 4 | Journal of Educational Computing Research | Journal | 3 |
| 5 | MDPI-Information | Journal | 3 |
| 6 | Computer Application Engineering Education | Journal | 2 |
| 7 | Educational Technology Research and Development | Journal | 2 |
| 8 | IEEE Transactions on Learning Technologies | Journal | 2 |
| 9 | Multimedia Tools and Applications | Journal | 2 |
| 10 | Simulation Gaming | Journal | 2 |
| 11 | ACM Transactions on Computing Education | Journal | 1 |
| 12 | Acta Didactica Napocensia | Journal | 1 |
| 13 | ARPN Journal of Engineering and Applied Sciences | Journal | 1 |
| 14 | Education Sciences | Journal | 1 |
| 15 | EMERALD INSIGHT | Journal | 1 |
| 16 | Entertainment Computing | Journal | 1 |
| 17 | Higher Education | Journal | 1 |
| 18 | IEEE Access | Journal | 1 |
| 19 | IEEE Latin America Transactions | Journal | 1 |
| 20 | IEEE-RITA | Journal | 1 |
| 21 | Informatics in Education | Journal | 1 |
| 22 | International Journal of Engineering Education | Journal | 1 |
| 23 | International Journal of Information Management | Journal | 1 |
| 24 | International Journal of Serious Games | Journal | 1 |
| 25 | International Journal of Technology Enhanced Learning | Journal | 1 |
| 26 | International Journal of Web Information Systems | Journal | 1 |
| 27 | Journal of Business Research | Journal | 1 |
| 28 | Journal of Systems Architecture | Journal | 1 |
| 29 | Jurnal Teknologi | Journal | 1 |
| 30 | MDPI-Computers | Journal | 1 |
| 31 | Revista | Journal | 1 |
| 32 | Universal Access in the Information Society | Journal | 1 |
| 33 | User Modeling and User-Adapted Interaction | Journal | 1 |
| 34 | 2017 40th International Convention on Information and Communication Technology, Electronics and Microelectronics (MIPRO) | Conference | 1 |
| 35 | 2017 6th IIAI International Congress on Advanced Applied Informatics | Conference | 1 |
| 36 | 2017 IEEE Symposium on Visual Languages and Human-Centric Computing (VL/HCC) | Conference | 1 |
| 37 | 2018 7th International Congress on Advanced Applied Informatics | Conference | 1 |
| 38 | 2020 IEEE Global Engineering Education Conference | Conference | 1 |
| 39 | 4th International Conference on Computing Sciences | Conference | 1 |

Continued on next page

Table 4: Publication Source (Continued)

| | | | |
|---|---|---|---|
| 40 | 6th Conference on Engineering Education (ICEED) | Conference | 1 |
| 41 | IEEE Global Engineering Education Conference (EDUCON) | Conference | 1 |
| 42 | Interactive Mobile Communication Technologies and Learning: Proceedings of the 11th IMCL Conference | Conference | 1 |
| 43 | Mobile Technologies and Applications for the Internet of Things: Proceedings of the 12th IMCL Conference | Conference | 1 |
| 44 | NordiCHI: Nordic Conference on Human-Computer Interaction | Conference | 1 |
| 45 | Procedia Computer Science | Conference | 1 |
| 46 | Proceedings of the 14th International Conference on the Foundations of Digital Games | Conference | 2 |
| 47 | Proceedings of the 15th International Academic MindTrek Conference on Envisioning Future Media Environments | Conference | 1 |
| 48 | Proceedings of the 15th International Conference on Computer Systems and Technologies - CompSysTech | Conference | 1 |
| 49 | Proceedings of the 18th ACM international conference on interaction design and children | Conference | 1 |
| 50 | Proceedings of the 2017 ACM Conference on International Computing Education Research | Conference | 1 |
| 51 | Proceedings of the 2019 ACM conference on innovation and technology in computer science education | Conference | 1 |
| 52 | Proceedings of the 2019 ACM Conference on International Computing Education Research | Conference | 1 |
| 53 | Proceedings of the 50th ACM technical symposium on computer science education | Conference | 1 |
| 54 | Proceedings of the 51st Hawaii International Conference on System Sciences. | Conference | 1 |
| 55 | Proceedings of the ACM Conference on Global Computing Education | Conference | 1 |
| 56 | Proceedings of the ACM on Human-Computer Interaction, 5(CHI PLAY), | Conference | 1 |
| 57 | Proceedings of the Sixth International Conference on Technological Ecosystems for Enhancing Multiculturality | Conference | 1 |
| 58 | Proceedings of the Working Group Reports on Innovation and Technology in Computer Science Education | Conference | 1 |
| 59 | Systems, Software and Services Process Improvement: 26th European Conference, EuroSPI 2019, Edinburgh, UK | Conference | 1 |
| 60 | World Conference on Educational Multimedia, Hypermedia Telecommunications | Conference | 1 |
| 61 | 8th International Symposium on Telecommunications (IST). | Conference | 1 |
| 62 | GHITALY@CHItaly. 1st Workshop on Games-Human Interaction | Conference | 1 |
| | **Total** | | **81** |

The data presented in Table 5 reveals that the maximum number of studies has been selected from highly recognized journals indexed in the Web of Science, and the rest of the studies picked good-rank conferences. Education and Information Technologies is at the top of the list, from which six studies were selected and then Interactive learning environment journal, from which three were selected. Similarly, Computer & Education, Journal of Educational

Computing Research, and MDPI-Information are the journals from which three studies were selected

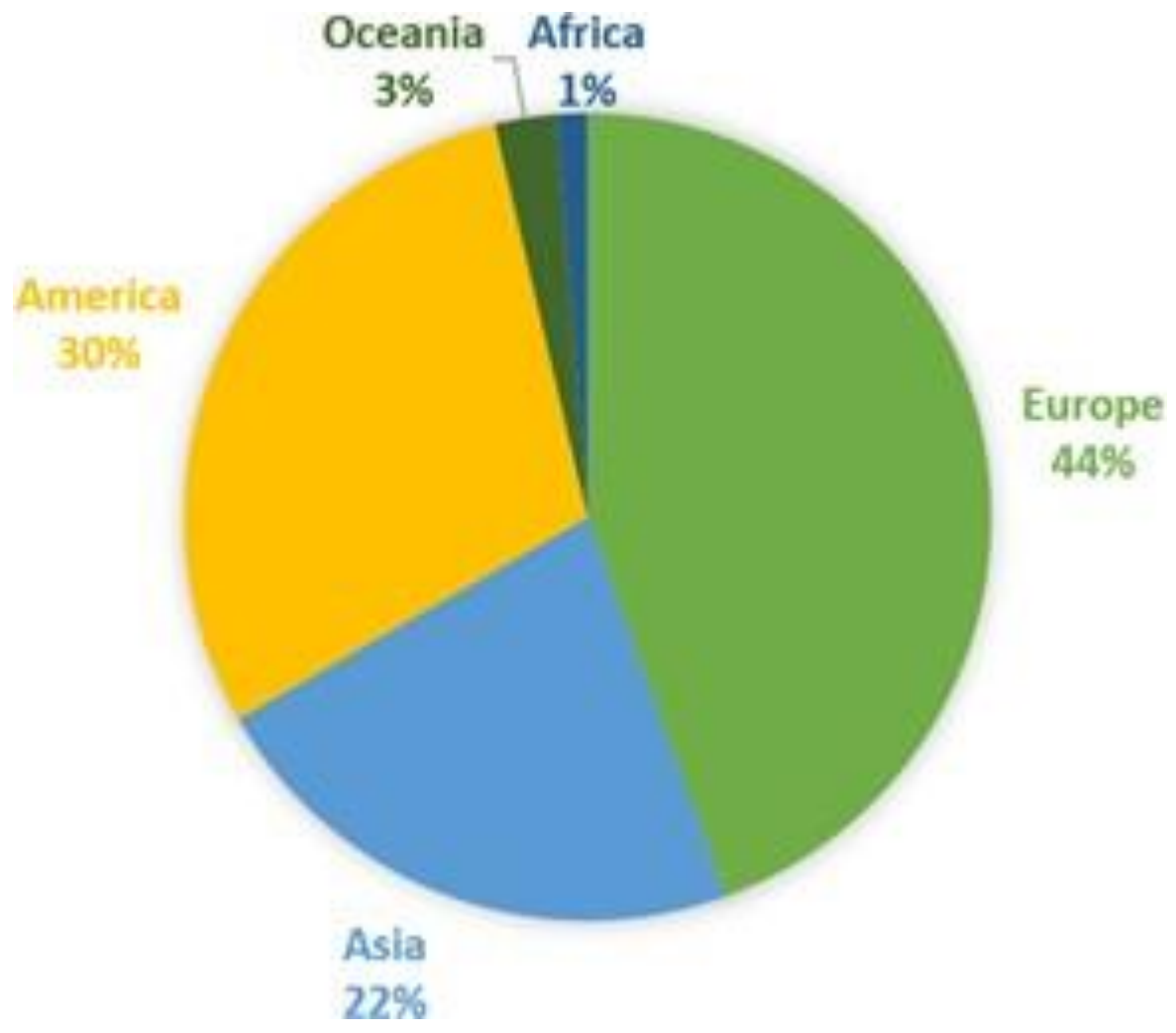

**Figure 2:** Publications by geographically

**RQ2: What criteria can be used to assess the quality of papers selected for review in the context of personalized gamified programming education, and how can these criteria be applied to the articles identified in RQ2 to ensure that only high-quality research is included in the dataset?**

The quality assessment (QA) score for each finalized study is awarded according to the criteria defined in section III, as shown in Table 5. Further, it shows the QA score range from 4-8, whereas a score less than 4 for the studies is discarded. Game-based programming language learning researchers may find this QA helpful in choosing related studies while addressing its usage and challenges. Articles published in Q1 journals mostly scored the highest, while studies scoring a total of 4 are from less recognized journals but relevant to the subject matter. 21 out of 81 scored highest, i.e. eight, which showed that all QA criteria were met by the studies, whereas 13 got the second highest score in the QA. Likewise, 12 out of 81 studies got the lowest score in the QA because they did not meet all the criteria. Overall classification results and QA of finalized studies have been presented in Table 6. Finalized studies have been classified based on five factors: empirical type, research type, and methodology.

Further, types of research have been categorized as Evaluation framework, Evaluation research, Solution proposal, and Review. All studies have empirically validated their results by performing statistical analysis, experiments, surveys, or case studies to increase their quality standards, awarded one score each. In category (c) of quality assessment criteria, only 11 out of 81 studies have not presented an empirical result that was awarded a zero score. In contrast, no study scored zero for category (d) of quality assessment criteria, but forty-five (45) studies got the lowest score in the same section. In addition, Table 8 presents the total studies that secure the highest to lowest scores accordingly.

Table 5: Quality assessment of the selected studies

| | | Classification | | | | | Quality Assessment | | | | |
|---|---|---|---|---|---|---|---|---|---|---|---|
| **Sr. No.** | **Ref.** | ***P. Chan-nel*** | ***P. Year*** | ***Research Type*** | ***Empirical Type*** | ***Methodology*** | ***(a)1*** | ***(b)2*** | ***(c)1*** | ***(d)4*** | ***Score*** |

Continued on next page

Table 5: Quality assessment of the selected studies (Continued)

| | | | | | | | | | | | |
|---|---|---|---|---|---|---|---|---|---|---|---|
| 1 | (Giannakoulas and Xinogalos, 2018) | Journal | 2018 | Evaluation Research | Survey | TAM model used | 1 | 2 | 1 | 4 | 8 |
| 2 | (Cabada et al., 2020) | Journal | 2020 | Evaluation Research | Experiment | TAM model used | 1 | 1 | 1 | 4 | 7 |
| 3 | (Malliarakis et al., 2017) | Journal | 2017 | Evaluation Research | Experiment and Survey | Evaluation Framework | 1 | 2 | 1 | 4 | 8 |
| 4 | (Papadakis, 2020) | Journal | 2020 | Evaluation Research | Experiment and Interview | Mix Method | 0 | 2 | 1 | 2 | 5 |
| 5 | (Topalli and Cagiltay, 2018) | Journal | 2018 | Solution Proposal | Experiment | Questionnaire | 1 | 2 | 1 | 4 | 8 |
| 6 | (Chang et al., 2020) | Journal | 2020 | Solution Proposal | Experiment and Survey | Questionnaire | 1 | 2 | 1 | 4 | 8 |
| 7 | (Jakoš and Verber, 2017) | Journal | 2017 | Evaluation Research | Experiment and Survey | Questionnaire | 1 | 2 | 1 | 4 | 8 |
| 8 | (Garneli and Chorianopoulos, 2018) | Journal | 2018 | Evaluation Research | Experiment | Interview and Observation | 1 | 1 | 1 | 4 | 7 |
| 9 | (Mathew et al., 2019) | Journal | 2019 | Solution Proposal | Survey | Interview | 1 | 1 | 1 | 3 | 6 |
| 10 | (Pellas and Vosinakis, 2018) | Journal | 2018 | Solution Proposal | Experiment | Questionnaire | 1 | 2 | 1 | 4 | 8 |
| 11 | (?) | Journal | 2021 | Solution Proposal | Experiment and Survey | Questionnaire and Interview | 1 | 2 | 1 | 4 | 8 |
| 12 | (Strawhacker and Bers, 2019) | Journal | 2019 | Evaluation Research | Experiment and Survey | Mix Method | 1 | 2 | 1 | 4 | 8 |
| 13 | (Hitchens and Tulloch, 2018) | Journal | 2018 | Solution Proposal | Survey | Questionnaire | 1 | 1 | 1 | 2 | 5 |
| 14 | (Syaifudin et al., 2020) | Journal | 2019 | Solution Proposal | N/A | N/A | 1 | 1 | 1 | 1 | 4 |
| 15 | (Marwan et al., 2019) | Conference | 2019 | Evaluation Research | Survey | Observation | 1 | 1 | 1 | 2 | 5 |

Continued on next page

Table 5: Quality assessment of the selected studies (Continued)

| | | | | | | | | | | | |
|---|---|---|---|---|---|---|---|---|---|---|---|
| 16 | (Maskeliu¯nas et al., 2020) | Journal | 2020 | Solution Proposal | Experiment and Survey | Questionnaire | 1 | 1 | 1 | 2 | 5 |
| 17 | (Duffany, 2017) | Journal | 2017 | Solution Proposal | N/A | Observation | 1 | 1 | 0 | 2 | 4 |
| 18 | (Seraj et al., 2018) | Conference | 2018 | Solution Proposal | N/A | Observation | 1 | 2 | 1 | 2 | 6 |
| 19 | (Figueiredo and García-Peñalvo, 2018) | Conference | 2018 | Evaluation Research | Review | Observation | 1 | 1 | 0 | 2 | 4 |
| 20 | (Krugel and Hubwieser, 2017) | Conference | 2017 | Solution Proposal | Survey | Questionnaire and Interview | 1 | 2 | 1 | 2 | 6 |
| 21 | (Skalka and Drlík, 2018) | Journal | 2018 | Solution Proposal | Review | Observation | 1 | 1 | 1 | 2 | 5 |
| 22 | (Nadolny et al., 2017) | Journal | 2017 | Evaluation Research | Survey | Questionnaire | 1 | 2 | 1 | 4 | 8 |
| 23 | (Hooshyar et al., 2018) | Journal | 2018 | Solution Proposal | Survey | Questionnaire and Interview | 1 | 2 | 1 | 4 | 8 |
| 24 | (Hausswolff, 2017) | Conference | 2017 | Evaluation Research | Survey | N/A | 1 | 1 | 0 | 2 | 4 |
| 25 | (Drosos et al., 2017) | Conference | 2017 | Solution Proposal | Survey | Questionnaire | 1 | 1 | 0 | 2 | 4 |
| 26 | (Bernik et al., 2017) | Conference | 2017 | Solution Proposal | Survey | Questionnaire | 1 | 1 | 1 | 2 | 5 |
| 27 | (Troiano et al., 2019) | Conference | 2019 | Solution Proposal | Survey | Questionnaire | 1 | 1 | 1 | 2 | 5 |
| 28 | (Malik et al., 2019) | Journal | 2019 | Solution Proposal | Survey | Questionnaire | 1 | 2 | 1 | 3 | 7 |
| 29 | (Devine et al., 2019) | Journal | 2019 | Evaluation Research | Survey | Questionnaire | 1 | 2 | 1 | 2 | 6 |
| 30 | (Yallihep and Kutlu, 2020) | Journal | 2020 | Evaluation Research | Experiment and Survey | Questionnaire | 1 | 2 | 1 | 3 | 7 |
| 31 | (Piedade et al., 2020) | Journal | 2020 | Evaluation Research | Mix Method | Questionnaire and Interview | 1 | 2 | 1 | 3 | 7 |
| 32 | (Luik et al., 2019) | Journal | 2019 | Evaluation Research | Mix Method | Questionnaire and Interview | 1 | 2 | 1 | 4 | 8 |
| 33 | (Simon et al., 2019) | Conference | 2019 | Evaluation Research | Survey | Questionnaire | 1 | 1 | 1 | 2 | 5 |

Continued on next page

Table 5: Quality assessment of the selected studies (Continued)

| | | | | | | | | | | | |
|---|---|---|---|---|---|---|---|---|---|---|---|
| 34 | (Martins et al., 2018) | Journal | 2018 | Solution Proposal | Experiment and Survey | Questionnaire | 1 | 2 | 1 | 3 | 7 |
| 35 | (Smith et al., 2019) | Conference | 2019 | Evaluation Research | Survey | Questionnaire | 1 | 2 | 1 | 2 | 6 |
| 36 | (Schez-Sobrino et al., 2020) | Journal | 2020 | Solution Proposal | Survey | Questionnaire | 1 | 2 | 1 | 3 | 7 |
| 37 | (Ivanović et al., 2017) | Journal | 2017 | Evaluation Research | Survey | Questionnaire | 1 | 2 | 1 | 4 | 8 |
| 38 | (Hellings and Haelermans, 2022) | Journal | 2022 | Evaluation Research | Experiment and Survey | Questionnaire | 1 | 2 | 1 | 3 | 7 |
| 39 | (Marwan et al., 2019) | Conference | 2019 | Evaluation Research | Survey | Questionnaire | 1 | 1 | 1 | 2 | 5 |
| 40 | (Laporte and Zaman, 2018) | Conference | 2018 | Evaluation Research | N/A | N/A | 1 | 1 | 0 | 2 | 4 |
| 41 | (Kumar and Sharma, 2019) | Conference | 2019 | Solution Proposal | N/A | N/A | 1 | 1 | 0 | 2 | 4 |
| 42 | (de Pontes et al., 2019) | Conference | 2019 | Solution Proposal | Survey | Questionnaire | 1 | 2 | 1 | 2 | 6 |
| 43 | (Paiva et al., 2020) | Journal | 2020 | Solution Proposal | Experiment and Survey | Questionnaire | 1 | 2 | 1 | 2 | 6 |
| 44 | (Wong and Yatim, 2018) | Conference | 2018 | Solution Proposal | Experiment and Survey | Questionnaire | 1 | 1 | 1 | 2 | 5 |
| 45 | (Gulec et al., 2019) | Conference | 2019 | Solution Proposal | Experiment and Survey | Questionnaire | 1 | 2 | 1 | 2 | 6 |
| 46 | (Tasadduq et al., 2021) | Journal | 2021 | Evaluation Research | Survey | Questionnaire | 1 | 2 | 1 | 3 | 7 |
| 47 | (Abbasi et al., 2021) | Journal | 2021 | Solution Proposal | Experiment and Survey | Questionnaire | 1 | 2 | 1 | 3 | 7 |
| 48 | (Zhu et al., 2019) | Conference | 2019 | Solution Proposal | Experiment and Survey | Questionnaire | 1 | 2 | 1 | 2 | 6 |
| 49 | (Sideris and Xinogalos, 2019) | Journal | 2019 | Solution Proposal | Experiment and Survey | Questionnaire | 1 | 2 | 1 | 4 | 8 |

Continued on next page

Table 5: Quality assessment of the selected studies (Continued)

| | | | | | | | | | | | |
|---|---|---|---|---|---|---|---|---|---|---|---|
| 50 | (Montes et al., 2021) | Journal | 2021 | Solution Proposal | Experiment and Survey | Questionnaire | 1 | 2 | 1 | 4 | 8 |
| 51 | (Xinogalos and Tryfou, 2021) | Journal | 2021 | Solution Proposal | Experiment and Survey | Questionnaire | 1 | 2 | 1 | 3 | 7 |
| 52 | (Toukiloglou and Xinogalos, 2022) | Journal | 2022 | Solution Proposal | Experiment and Survey | Questionnaire | 1 | 2 | 1 | 4 | 8 |
| 53 | (Daungcharone et al., 2017) | Conference | 2017 | Solution Proposal | Experiment and Survey | Questionnaire | 1 | 2 | 1 | 2 | 6 |
| 54 | (Carreño-León et al., 2019) | Conference | 2019 | Solution Proposal | Experiment and Survey | Questionnaire | 1 | 2 | 1 | 2 | 6 |
| 55 | (Jemmali et al., 2019) | Conference | 2019 | Solution Proposal | N/A | N/A | 1 | 1 | 0 | 2 | 4 |
| 56 | (Khaleel et al., 2019) | Journal | 2019 | Solution Proposal | Experiment and Survey | Questionnaire | 1 | 2 | 1 | 2 | 6 |
| 57 | (Moreno and Pineda, 2018) | Journal | 2018 | Evaluation Research | N/A | N/A | 1 | 1 | 0 | 2 | 4 |
| 58 | (Sanmugam et al., 2014) | Conference | 2014 | Evaluation Research | N/A | N/A | 1 | 1 | 1 | 2 | 5 |
| 59 | (Paiva et al., 2020) | Conference | 2020 | Evaluation Research | Survey | Questionnaire | 1 | 1 | 1 | 2 | 5 |
| 60 | (Queirós, 2019) | Journal | 2019 | Solution Proposal | Experiment and Survey | Questionnaire | 1 | 1 | 1 | 2 | 5 |
| 61 | (Azmi et al., 2015) | Journal | 2015 | Evaluation Research | Survey | Questionnaire | 1 | 1 | 1 | 2 | 5 |
| 62 | (Bennani et al., 2020) | Journal | 2020 | Evaluation Research | Survey | N/A | 1 | 1 | 1 | 2 | 5 |
| 63 | (Challco et al., 2015) | Journal | 2015 | Solution Proposal | Experiment and Survey | Questionnaire | 1 | 2 | 1 | 4 | 8 |
| 64 | (De-Marcos et al., 2016b) | Journal | 2016 | Solution Proposal | Experiment and Survey | Questionnaire | 1 | 2 | 1 | 4 | 8 |
| 65 | (Johnson et al., 2016) | Conference | 2016 | Evaluation Research | Survey | Questionnaire | 1 | 1 | 1 | 2 | 5 |

Continued on next page

Table 5: Quality assessment of the selected studies (Continued)

| | | | | | | | | | | | |
|---|---|---|---|---|---|---|---|---|---|---|---|
| 66 | (Hassan et al., 2021) | Journal | 2021 | Evaluation Research | Mix Method | Questionnaire and Interviews | 1 | 2 | 1 | 4 | 8 |
| 67 | (Katan and Anstead, 2020) | Conference | 2020 | Evaluation Research | Survey | Questionnaire | 1 | 1 | 1 | 2 | 5 |
| 68 | (Kiraly and Balla, 2020) | Journal | 2020 | Evaluation Research | Survey | Questionnaire | 1 | 1 | 1 | 2 | 5 |
| 69 | (Knutas et al., 2014) | Conference | 2014 | Evaluation Research | Survey | Questionnaire | 1 | 1 | 1 | 2 | 5 |
| 70 | (Knutas et al., 2017) | Conference | 2017 | Evaluation Research | Survey | Questionnaire | 1 | 1 | 1 | 2 | 5 |
| 71 | (Knutas et al., 2019) | Journal | 2019 | Solution Proposal | Experiment and Survey | Questionnaire | 1 | 2 | 1 | 4 | 8 |
| 72 | (Marín et al., 2018) | Journal | 2019 | Solution Proposal | Experiment and Survey | Questionnaire | 1 | 2 | 1 | 4 | 8 |
| 73 | (Mullins and Sabherwal, 2018) | Conference | 2018 | Evaluation Research | Survey | Questionnaire | 1 | 1 | 0 | 2 | 4 |
| 74 | (Mullins and Sabherwal, 2020) | Journal | 2020 | Evaluation Research | Survey | N/A | 1 | 1 | 0 | 2 | 4 |
| 75 | (de Marcos Ortega et al., 2017) | Journal | 2017 | Evaluation Research | Survey | Questionnaire | 1 | 1 | 1 | 2 | 5 |
| 76 | (Rodrigues et al., 2021) | Conference | 2021 | Evaluation Research | Survey | Questionnaire | 1 | 1 | 1 | 2 | 5 |
| 77 | (Rodrigues et al., 2022) | Journal | 2022 | Solution Proposal | Experiment and Survey | Questionnaire | 1 | 2 | 1 | 4 | 8 |
| 78 | (Rojas-López et al., 2019) | Journal | 2019 | Solution Proposal | Experiment and Survey | Questionnaire | 1 | 1 | 1 | 3 | 6 |
| 79 | (Roosta et al., 2016) | Conference | 2016 | Evaluation Research | Survey | N/A | 1 | 1 | 0 | 2 | 4 |
| 80 | (Santos et al., 2021) | Journal | 2021 | Solution Proposal | Experiment and Survey | Questionnaire | 1 | 2 | 1 | 3 | 7 |
| 81 | (Toda et al., 2019) | Journal | 2019 | Evaluation Research | Survey | Questionnaire | 1 | 2 | 1 | 3 | 7 |

Table 6: Accumulative Quality Assessment Score

| References | Score | Total |
|---|---|---|
| ((Challco et al., 2015) (Chang et al., 2020) (De-Marcos et al., 2016b) (Giannakoulas and Xinogalos, 2018) (Hassan et al., 2021) (Hooshyar et al., 2018) (Ivanović et al., 2017) (Jakoš and Verber, 2017) (Knutas et al., 2019) (Luik et al., 2019) (Malliarakis et al., 2017) (Marín et al., 2018) (Montes et al., 2021) (Nadolny et al., 2017) (Pellas and Vosinakis, 2018) (Rodrigues et al., 2022) (Sideris and Xinogalos, 2019) (Strawhacker and Bers, 2019) (Topalli and Cagiltay, 2018) (Toukiloglou and Xinogalos, 2022) (**?**) | 8 | 21 |
| (Abbasi et al., 2021) (Garneli and Chorianopoulos, 2018) (Hellings and Haelermans, 2022) (Malik et al., 2019) (Martins et al., 2018) (Piedade et al., 2020) (Santos et al., 2021) (Schez-Sobrino et al., 2020) (Toda et al., 2019) (Tasadduq et al., 2021) (Xinogalos and Tryfou, 2021) (Yallihep and Kutlu, 2020) (Cabada et al., 2020) | 7 | 13 |
| (Carreño-León et al., 2019) (Daungcharone et al., 2017) (de Pontes et al., 2019) (Devine et al., 2019) (Gulec et al., 2019) (Khaleel et al., 2019) (Krugel and Hubwieser, 2017) (Mathew et al., 2019) (Paiva et al., 2020) (Rojas-López et al., 2019) (Seraj et al., 2018) (Smith et al., 2019) (Zhu et al., 2019) | 6 | 13 |
| (Azmi et al., 2015) (Bennani et al., 2020) (Bernik et al., 2017) (de Marcos Ortega et al., 2017) (Johnson et al., 2016) (Hitchens and Tulloch, 2018) (Katan and Anstead, 2020) (Kiraly and Balla, 2020) (Knutas et al., 2014) (Knutas et al., 2017) (Luxton-Reilly et al., 2019) (Marwan et al., 2019) (Maskeliu¯nas et al., 2020) (Pankiewicz, 2020) (Papadakis, 2020) (Queirós, 2019) (Rodrigues et al., 2021) (Sanmugam et al., 2014) (Skalka and Drlík, 2018) (Troiano et al., 2019) (Wong and Yatim, 2018) | 5 | 22 |
| (Drosos et al., 2017) (Duffany, 2017) (Figueiredo and García-Peñalvo, 2018) (Jemmali et al., 2019) (Kumar and Sharma, 2019) (Laporte and Zaman, 2018) (Moreno and Pineda, 2018) (Mullins and Sabherwal, 2018) (Mullins and Sabherwal, 2020) (Roosta et al., 2016) (Syaifudin et al., 2020) (Hausswolff, 2017) | 4 | 12 |

**RQ3: What are the trends and best practices for implementing personalized gamification frameworks in programming education, and how do these frameworks differ in design and customization?** Personalized gamified programming education has emerged as an innovative and engaging approach to enhancing students' learning experiences. However, designing an effective personalized gamified programming education system requires a deep understanding of the relevant theories and frameworks/conceptual models used in this context. This section aims to identify the frameworks/conceptual models that have been applied to personalized gamified programming education with respect to students' cognition research and explore the relationships between them. This section will provide a comprehensive overview of the theoretical foundations underpinning personalized gamified programming education, which can serve as a valuable resource for researchers and educators in this field. This research question will explore using adopted and custom frameworks in gamification for programming education. In our analysis, several papers did not mention any specific gamification framework. We referred to them as 'Not specified' (NS).

Table 7 summarizes the studies that used each gamification framework for programming language education interventions. The "Adopted Frameworks" category includes previously developed and used in other contexts, while the "Custom Frameworks" category includes frameworks specifically designed for the intervention. The "Not Specified (NS)" category includes studies that did not explicitly mention using any gamification framework.

Table 7: Summary of Gamification Frameworks Used in Programming Language Education

| Gamification Framework | | Number of Studies |
|---|---|---|
| Adopted Frameworks | Technology Acceptance Model (TAM) | 1 |

Continued on next page

Table 7: Summary of Gamification Frameworks Used in Programming Language Education (Continued)

| | | |
|---|---|---|
| | Attention, Relevance, Confidence, and Satisfaction (ARCS) | 2 |
| | TETEM | 1 |
| Custom Frameworks | | 14 |
| Not Specified (NS) | | 15 |

**Adopted Gamification Frameworks:** Previous research has used various gamification frameworks such as the Technology Acceptance Model (TAM), the ARCS model, and the Turkish Educational Technology Evaluation Model (TETEM). The TAM framework investigates learners' acceptance of gamification elements, while the ARCS model aims to motivate learners by drawing their attention to the material, emphasizing its relevance, building their confidence, and providing satisfaction and rewards. TETEM is a framework used to evaluate educational technologies in the Turkish context and has been used in several research studies. Despite being popular, only two papers used ARCS in the context of gamification for programming language education. Among the 36 papers reviewed, only 2 studies used frameworks such as TAM and TETEM, while ARCS was used in 2 papers. For instance, (Cabada et al., 2020) employed the TAM framework to evaluate students' perceptions of automated programming hints. Another study by (Maskeliu¯nas et al., 2020) used both TAM and TETEM to assess the effectiveness of an interactive mobile game for learning programming. (Khaleel et al., 2019) used the ARCS model to guide the design of a gamified learning system to improve learning outcomes in a programming language website.

**Custom Gamification Frameworks:** Several studies have explored gamification techniques to enhance programming education. Custom gamification frameworks provide greater control over the design and implementation of gamification techniques, but their development can be resource-intensive and require high technical expertise. Some studies have used custom frameworks, such as CMX (Malliarakis et al., 2017), a microlearning-based mobile application (Skalka and Drlík, 2018), and a game-based Bayesian intelligent tutoring system (Hooshyar et al., 2018). (Kumar and Sharma, 2019) demonstrated improved student engagement and learning outcomes through a gamified approach that used Bayesian networks as a decision-making tool. (de Pontes et al., 2019) and (Paiva et al., 2020) implemented frameworks that provided platforms for programming exercises and assessments, both of which improved student engagement and learning outcomes. (Tasadduq et al., 2021) utilized a custom framework to evaluate the impact of gamification on students with a background in rote learning who are learning computer programming. (Abbasi et al., 2021) investigated the effectiveness of serious games in enhancing student's learning performance and motivation using a custom framework. Several papers investigate the use of serious games and gamification in programming education. (Zhu et al., 2019) utilize a serious game framework to teach parallel programming, while (Sideris and Xinogalos, 2019) present a framework that teaches programming concepts through a 2D platform game. (Xinogalos and Tryfou, 2021) discuss using Greenfoot as a tool for creating serious games for programming education, and (Daungcharone et al., 2017) describe a gaming framework that employs a digital game as a compiler to motivate C programming language learning in higher education. (Carreño-León et al., 2019) introduce a gaming framework that uses gamification techniques to enhance problem-solving skills in programming education, incorporating tailored challenges, a scoring system, and a feedback mechanism to increase student engagement and motivation. Lastly, (Marín et al., 2018) conduct an empirical study on the effectiveness of gamification techniques in programming courses, incorporating social gamification elements.

**No Mention of Gamification Frameworks:** Approximately 15 papers reviewed did not explicitly mention using gamification frameworks for programming education. (Topalli and Cagiltay, 2018) proposed using a problem-based learning approach to encourage collaborative game development without a specific framework. (Mathew et al., 2019) utilized an educational game called PROSOLVE, incorporating problem-based learning and gamification techniques. (Hitchens and Tulloch, 2018) designed a gamification approach for classroom instruction, integrating game elements such as rewards, feedback, and progress tracking, but did not mention a specific framework. The authors use game design principles such as immediate feedback and gradual increase in difficulty levels to design activities that include badges, points, and leaderboards. Various studies have explored different approaches to programming language education. While some have used specific gamification frameworks, others have not. For instance, (Syaifudin et al., 2020) proposed an Android Programming Learning Assistant System (APLAS) to help students learn basic Android application development, while (Marwan et al., 2019) used a problem-based learning approach and automated programming hints to improve students' performance in programming. Additionally, some studies have focused on active learning techniques such as pair programming and think-pair-share, like the work of (Duffany, 2017), while others have explored the use of educational robotics, such as (Piedade et al., 2020). Furthermore, some studies, like (Luik et al., 2019), did not discuss any explicit gaming aspect. However, incorporating gaming aspects such as challenges, points, levels, and feedback can enhance the learning experience in programming language education, as suggested by various studies.

**Game-Based Learning (GBL):** Gamification frameworks are commonly used in programming language education interventions to enhance learner engagement and motivation. However, game-based learning (GBL), which involves using games for learning, is another approach rather than applying gamification elements to a non-game context. In programming language education, GBL typically involves designing games or game-like activities that require learners to apply programming concepts to progress or succeed. One example of GBL in programming language education is a mobile application developed by (Chang et al., 2020), which incorporated game elements such as points, badges, and leaderboards to enhance learner engagement and motivation.

**Validation of framework:** Based on our analysis of available information, we found that some gamification frameworks have been validated, often through Structural Equation Modeling (SEM) or questionnaires. However, the success of a framework depends on various factors and careful consideration, and ongoing evaluation is necessary when adopting or customizing a framework for a specific purpose. It is also important to note that many studies reviewed did not explicitly mention a framework, making it difficult to compare the effectiveness of different interventions. Table 8 summarizes the references and types of gamification frameworks used in programming language education interventions.

Our literature review identified effective practices and trends in personalized gamification frameworks for programming education. Personalization increased student motivation and engagement, and four categories of papers were identified: those adopting established frameworks, customized frameworks, game elements, and those combining gamification and game-based learning. Educators and instructional designers can use these insights to create effective and engaging learning experiences. However, challenges such as clear goal setting and potential distraction from learning objectives were also identified. Further research is needed to examine the effectiveness of different personalized gamification frameworks in different contexts, and empirical studies are needed to evaluate their effectiveness in real-world settings.

**RQ4: How can personalized gamification frameworks in programming education be mapped to the different cognitive levels of Bloom's taxonomy?**

This research question explores the effective alignment of frameworks with different cognitive levels of Bloom's taxonomy to understand how personalized gamification can enhance learning outcomes in programming education. This involves identifying the levels of learning in Bloom's taxonomy and analyzing how they relate to gamification framework design.

Bloom's Taxonomy categorizes educational goals into different levels of cognitive complexity. These levels range from lower-order thinking skills, such as remembering and understanding, to higher-order thinking skills, such as synthesizing and evaluating complex information and ideas. The categories of Bloom's Taxonomy can be divided into low-level and high-level thinking skills.

**Low-level thinking skills (LL):**

- *Remembering:* recalling facts, information, or procedures.
- *Understanding:* comprehending the meaning of information, including identifying patterns and relationships. The lower levels of the taxonomy (knowledge, comprehension, and application) involve basic cognitive processes such as memorization, understanding, and application of information.

**High-level thinking skills (HL):**

- Applying: using knowledge and skills to solve problems or complete tasks in new situations.
- Analyzing: breaking down complex information into smaller parts to better understand it.

**Table 19**
References and Types of Gamification Frameworks Used in Programming Language Education

| References | Adopted/ Custom | Based on |
|---|---|---|
| (Cabada et al., 2020) | Adopted | TAM |
| (Maskeliu¯nas et al., 2020) | Adopted | TAM and TETEM |
| (Khaleel et al., 2019) | Adopted | ARCS |
| (Malliarakis et al., 2017) | Custom | constructivist learning theory |
| (Skalka and Drlík, 2018), (Kumar and Sharma, 2019), (de Pontes et al., 2019), (Tasadduq et al., 2021) , (Zhu et al., 2019), (Daungcharone et al., 2017), (Carreño-León et al., 2019) , (Marín et al., 2018) | Custom | No mention |
| (Hooshyar et al., 2018) | Custom | Bayesian network based |
| (Paiva et al., 2020) | Custom | Asura |
| (Abbasi et al., 2021) | Custom | SG model |
| (Sideris and Xinogalos, 2019) | Custom | Educational Games Design Model proposed by (Ibrahim and Jaafar, 2009) |
| (Xinogalos and Tryfou, 2021) | Custom | Serious Game Design Assessment (SGDA) Framework was created by Mitgutsch and Alvarado |
| (Topalli and Cagiltay, 2018) (Hitchens and Tulloch, 2018), (Syaifudin et al., 2020), (Duffany, 2017) (Figueiredo and García-Peñalvo, 2018) (Skalka and Drlík, 2018), (Malik et al., 2019), (Piedade et al., 2020), (Luik et al., 2019), (Simon et al., 2019), (Hellings and Haelermans, 2022), (De-Marcos et al., 2016a), (Bernik et al., 2017) | NS | - |
| (Mathew et al., 2019) | NS | They incorporate elements of problem-based learning |
| (Marwan et al., 2019) | NS | Does use a problem-based learning |
| (Chang et al., 2020) | - | They used the GBL design model proposed by (Shi and Shih, 2015) |

- Evaluating: making judgments about the value or quality of information or ideas.
- Creating: combining knowledge and skills to create something new or original

cognitive processes such as breaking down information into parts, combining ideas to form a new whole, and making judgments about the value or quality of information. Our paper's analysis classified them into high-level (HL) and low-level (LL) Bloom taxonomy categories. Out of the 34 papers analyzed, 26 aligned with high-level thinking, 1 with low-level thinking, 6 aligned with both high and low-level thinking, and 1 did not align with either category. For papers where the taxonomy was not clearly stated, we inferred the level of thinking based on the study's outcome. Table 9 presents the references of the 34 papers analyzed in this study and their alignment with Bloom's taxonomy. The papers were categorized as high level (HL), low level (LL), both high and low level (HL and LL), or no alignment based on their focus on either higher-order thinking skills or foundational knowledge.

Our literature review found that gamification can increase student engagement and motivation in programming education, especially when using Bloom's taxonomy to design activities that enhance cognitive complexity. The effectiveness of gamification in achieving learning outcomes depends on factors such as the specific outcomes being targeted and the design of the activity. Our findings suggest that gamification and Bloom's taxonomy can positively impact motivation, cognitive complexity, and learning outcomes, providing important insights for educators and instructional designers.

**Table 9**
Bloom taxonomy alignment

| References | Bloom Taxonomy – level |
|---|---|
| (Malliarakis et al., 2017) , (Topalli and Cagiltay, 2018) , (Chang et al., 2020) , (Mathew et al., 2019), (Hitchens and Tulloch, 2018), (Syaifudin et al., 2020), (Marwan et al., 2019), (Duffany, 2017), (Figueiredo and García-Peñalvo, 2018), (Skalka and Drlík, 2018), (Hooshyar et al., 2018), (Malik et al., 2019), (Piedade et al., 2020), (Luik et al., 2019), (Wong and Yatim, 2018), (Tasadduq et al., 2021), (Abbasi et al., 2021), (Zhu et al., 2019), (Sideris and Xinogalos, 2019), (Xinogalos and Tryfou, 2021), (Daungcharone et al., 2017) , (Carreño-León et al., 2019) , (Khaleel et al., 2019) , (de Marcos Ortega et al., 2017), (Marín et al., 2018) | HL |
| (Maskeliu¯nas et al., 2020) | LL |
| (Cabada et al., 2020), (Hellings and Haelermans, 2022), (Kumar and Sharma, 2019), (de Pontes et al., 2019) , (Paiva et al., 2020), (Bernik et al., 2017) | LL and HL |
| (Simon et al., 2019) | No alignment with Bloom |

### Gamification Aspect and Blooms Taxonomy

This subsection explores the relationship between gamification activities in programming education and Bloom's Taxonomy. Specifically, we examine how gamification activities align with different levels of cognitive complexity and promote higher-order thinking skills. By analyzing the gamification aspect with Bloom's Taxonomy, this subsection provides important insights into designing effective and engaging learning activities that promote higher-order thinking skills and support programming education. Specifically, we focus on several key gamification aspects, including:

1) *Intrinsic motivation* is the internal drive to engage in a task or activity because it is personally rewarding or satisfying. This type of motivation can be aligned with the higher levels of Bloom's taxonomy, specifically the levels of evaluating and creating. It can help learners stay engaged and motivated during these challenging tasks, as they derive enjoyment and satisfaction from the learning process.
2) *Extrinsic motivation* is driven by external rewards or consequences and aligns with the lower levels of Bloom's taxonomy. It helps learners stay motivated by providing a goal to achieve or avoid negative consequences.
3) *Performance gain* aligns with Bloom's taxonomy levels of applying and analyzing knowledge and skills, where learners are expected to use their knowledge to solve problems and complete tasks effectively. It is measured by how well learners can apply acquired knowledge and skills in real-world situations, such as writing functional code in a programming course.
4) *Attention and engagement* are important prerequisites for effective memory encoding and retrieval, and they can be aligned with Bloom's first level of remembering. When learners are engaged and attentive, they are more likely to process information deeply and form strong memory representations, which can be retrieved later when needed.
5) *Feedback and Assessment:* Gamification provides learners constructive feedback on their progress and performance. It aligns with Bloom's taxonomy's higher levels. Feedback helps learners identify errors or gaps in their understanding, while assessment evaluates learners' ability to judge the value, quality, or effectiveness of ideas, products, or solutions.
6) *Collaboration and Social Learning:* Gamification facilitates collaboration and social learning among learners, supporting the development of higher-level thinking skills.
7) *Creativity and Innovation:* It aligns with Bloom's taxonomy by promoting creativity and innovation at the highest level of cognitive taxonomy. It encourages learners to use their imagination and problem-solving skills through engaging and challenging activities. In analyzing 34 papers, we found that intrinsic motivation in gamification aligns with Bloom's creating and evaluating categories, while extrinsic motivation aligns with applying and analyzing. However, gamification literature has no clear distinction between the two types of motivation. The relationship between motivation and Bloom's taxonomy may vary based on gamification's specific context and application. Though studies have taken different approaches, aligning gamification with Bloom's categories can offer useful insights for incorporating it into educational settings.

**RQ5. What tools and software applications are developed based on personalized gamification frameworks in programming education, and how are these tools tailored to specific programming languages and concepts?** This section identified several tools and software applications developed based on personalized gamification frameworks in programming education. The results are summarized in Table 10. These tools incorporate game elements and mechanics into programming tasks to improve student motivation, engagement, and learning outcomes. The frameworks used to develop these tools vary. Despite differences in frameworks, the tools share common features such as badges, points, leaderboards, and rewards to incentivize student performance. Most tools are tailored to specific programming languages and concepts and provide personalized feedback and adaptive challenges to meet individual learner needs. Our review suggests that personalized gamification can effectively enhance student motivation, engagement, and learning outcomes in programming education. The development of tailored tools and software applications that align with specific programming languages and concepts can further enhance the effectiveness of gamification in programming education. However, more research is needed to evaluate the long-term effects of these tools on student learning outcomes and to identify best practices for designing and implementing gamified programming education tools.

Table 10: Tool and applications

| Reference | Topic/Concept | Gamification Framework | Tool/Software Application |
|---|---|---|---|
| (Cabada et al., 2020) | Algorithm and code construction | Adopted | EasyLogic |
| (Malliarakis et al., 2017) | General programming concepts | Custom | CMX environment |
| (Topalli and Cagiltay, 2018) | Introduction to programming course | NS | Gamification of exercises - physical |
| (Chang et al., 2020) | Introductory course | GBL | Programmer Adventure land |
| (Mathew et al., 2019) | Introductory programming course | NS | PROSOLVE game based on pseudo-code technique. |
| (Hitchens and Tulloch, 2018) | No topic mentioned in the paper | NS | Classroom activities and associated software were designed and implemented |
| (Syaifudin et al., 2020) | Java | NS | Android Programming Learning Assistance System, namely APLAS. |
| (Marwan et al., 2019) | General programming concepts | NS | They used iSnap and added hints to it. |
| (Maskeliu¯nas et al., 2020) | Javascript | NS | They developed a game. No name is mentioned. |
| (Duffany, 2017) | Visual Basic | NS | The classroom activities were designed to support active learning |
| (Figueiredo and García-Peñalvo, 2018) | To program mobile | NS | Block-based Enduser |

Continued on next page

Table 10: Tool and applications (Continued)

| | | | |
|---|---|---|---|
| | robots, microcontrollers and smart environments | | programming tool |
| (Skalka and Drlík, 2018) | Introduction to computational thinking and object-oriented concepts | NS | MOOC called LOOP (Learning Object-oriented Programming) |
| (Skalka and Drlík, 2018) | Programming concept | Custom | Only framework is proposed |
| (Hooshyar et al., 2018) | Introduction to programming course | Custom | Online Game-based Bayesian Intelligent Tutoring System (OGITS) |
| (Bernik et al., 2017) | Batch and Stack | NS | They used Moodle with Gamification features |
| (Malik et al., 2019) | Introductory programming (IP) courses | NS | PROBSOL |
| (Piedade et al., 2020) | Programming fundamentals | NS | Not mentioned |
| (Kumar and Sharma, 2019) | Programming concepts | Custom | Development of ProLounge (Programming Lounge) - an online learning application. |
| (de Pontes et al., 2019) | Introductory Programming course | Custom | They designed a game |
| (Paiva et al., 2020) | Game-based programming challenges (Java) | Custom | ASURA |
| (Wong and Yatim, 2018) | OOP | Adopted | Odyssey of Phoenix |
| (Tasadduq et al., 2021) | C | Custom | CYourWay |
| (Abbasi et al., 2021) | OOP | Custom | A 2D game named as Object Oriented serious game (OOsg) |
| (Zhu et al., 2019) | Concurrent and parallel programming (CPP) skills | Custom | Parallel |
| (Sideris and Xinogalos, 2019) | programming concepts using Python | Custom | PY-RATE ADVENTURES |
| (Xinogalos and Tryfou, 2021) | OOP | Custom | Game of Code: Lost in Javaland |
| (Daungcharone et al., 2017) | C | Custom | a digital game named CPGame |
| (Carreño-León et al., 2019) | Introductory programming course | Custom | No tool was developed |
| (Khaleel et al., 2019) | OOP (Java) | Adopted | The gamified website was developed |
| (Marín et al., 2018) | C | Custom | A gamified platform, namely UDPiler |

**RQ6: What are the common processes, tools, and instruments utilized for evaluating applications based on personalized gamified programming education? What evaluation measures are employed to assess applications**

**Table 11**
Methodology adopted by the studies

| Source Ref. | MethodoTlotgayl |
|---|---|
| (Abbasi et al., 2021) (Carreño-León et al., 2019) (Chang et al., 2020) (Daungcharone et al., 2017) (Giannakoulas and Xinogalos, 2018) (Garneli and Chorianopoulos, 2018) (Gulec et al., 2019) (Hellings and Haelermans, 2022) (Ivanović et al., 2017) (Jakoš and Verber, 2017) (Khaleel et al., 2019) (Kumar and Sharma, 2019) (Malliarakis et al., 2017) (Marín et al., 2018) (Marín et al., 2018) (Marwan et al., 2019) (Mathew et al., 2019) (Montes et al., 2021) (Moreno and Pineda, 2018) (Paiva et al., 2020) (Pankiewicz, 2020) (Papadakis, 2020) (Pellas and Vosinakis, 2018) (Schez-Sobrino et al., 2020) (Sideris and Xinogalos, 2019) (Smith et al., 2019) (Strawhacker and Bers, 2019) (Tasadduq et al., 2021) (Topalli and Cagiltay, 2018) (Toukiloglou and Xinogalos, 2022) (Troiano et al., 2019) (?) (Wong and Yatim, 2018) (Xinogalos and Tryfou, 2021) (Yallihep and Kutlu, 2020) (Cabada et al., 2020) (Zhu et al., 2019) | Quantitat3i7ve |
| (Hooshyar et al., 2018) (Krugel and Hubwieser, 2017) (Marwan et al., 2019) | Qualitativ3e |

**from various viewpoints, such as teaching, learning, and technical perspectives?**

The methodology refers to the fundamental techniques or methods used to identify, collect, retrieve, and interpret information on the topic (Paul, 2000). This research question posed to examine the tools and evaluation methodologies by the selected studies is presented in Tables 11 and 12.

Table 11 shows that 37 out of 81 studies used quantitative research methodology by asking questions from participants in a questionnaire/survey. This research question explored the tools and evaluation methodologies used by the selected studies. According to the evaluation tools, 33 studies, which are the majority, used Statistical Package for Social Science (SPSS) to evaluate the data accordingly, whereas nominal studies used Microsoft Excel. The methodology and tools used by the selected studies are presented in Table 12.

In this section, the evaluation measures terminologies are described from the selected studies:

- Descriptive: Descriptive statistics are short informative coefficients that describe a specific data collection, which might represent the full population or a subset of a population (Hayes). The descriptive results presented by (Abbasi et al., 2021), (Giannakoulas and Xinogalos, 2018), (Gulec et al., 2019), (Malliarakis et al., 2017), (Marín et al., 2018), (Martins et al., 2018), (Marwan et al., 2019), (Mathew et al., 2019), (Montes et al., 2021) (Paiva et al., 2020), (Schez-Sobrino et al., 2020), (Sideris and Xinogalos, 2019), (Smith et al., 2019), (Tasadduq et al., 2021), (Troiano et al., 2019), (Wong and Yatim, 2018), (Xinogalos and Tryfou, 2021), (Yallihep and Kutlu, 2020), (Cabada et al., 2020), and (Zhu et al., 2019) Frequencies: A frequency distribution is a visual or tabular display that shows the number of occurrences over a specific period (Young) calculated by (Carreño-León et al., 2019) and (Toukiloglou and Xinogalos, 2022).

- Mean: It is the average of the data set (adding all the numbers then dividing by its total point) (?)was calculated by (Daungcharone et al., 2017), (Jakoš and Verber, 2017), (Khaleel et al., 2019), (Moreno and Pineda, 2018), (Papadakis, 2020), (Kalogiannakis and Papadakis, 2019), and (Pellas and Vosinakis, 2018).

- Standard Deviation (SD): SD is the square root of the variance, which measures how to spread out a set of numbers compared to its mean (Hargrave). It was calculated by (Daungcharone et al., 2017), (Jakoš and Verber, 2017), (Khaleel et al., 2019), (Malliarakis et al., 2017), (Moreno and Pineda, 2018), (Papadakis, 2020), and (Pellas and Vosinakis, 2018).

- T-test: The independent t-test compares two collections of data, each of which is centred on a constant value, to determine whether or not there is statistical significance between them (e.g., interval or ratio) (Anova) was calculated by (Abbasi et al., 2021), (Chang et al., 2020), (Khaleel et al., 2019), (Tasadduq et al., 2021), (Topalli and Cagiltay, 2018), (Wong and Yatim, 2018), (Yallihep and Kutlu, 2020), and (Cabada et al., 2020) Analysis of Variance: The statistical technique known as analysis of variance (ANOVA) is used to compare multiple groups using a dependent variable that has two or more discrete categories (Anova), which were calculated by (Jakoš and Verber, 2017), and (Wong and Yatim, 2018).

- Analysis of covariance: A continuous variable is added to the variables of interest in an analysis of covariance (ANCOVA) (i.e., the dependent and independent variable) as means for control (StatisticsSolutions2013).
- Multivariate Analysis of Variance: The goal of multivariate analysis of variance (MANOVA), which is similar to ANOVA, is to examine differences between groups by using two or more dependent variables as opposed to one metric dependent variable (Anova) calculated by (Daungcharone et al., 2017).
- Wilcoxon signed-rank test (z): Two paired groups can be compared using the nonparametric Wilcoxon test, which can be either the rank sum test or the signed-rank test. The tests effectively compute the difference between groups of pairings and examine it to see if it is statistically significant (Hayes), calculated by (Garneli and Chorianopoulos, 2018).
- Mann–Whitney U test: When the dependent variable is ordinal or continuous but not normally distributed, the Mann-Whitney U test is used to examine the differences between two groups (Anova) calculated by (Garneli and Chorianopoulos, 2018), (Paiva et al., 2020), Pankiewicz (2020), (Pellas and Vosinakis, 2018), and (Tasadduq et al., 2021).
- Kruskal–Wallis: The medians of three or more independent groups are compared using the Kruskal-Wallis test to evaluate whether or not there is a statistically significant difference (Zach) calculated by (Strawhacker and Bers, 2019), and (Toukiloglou and Xinogalos, 2022).
- Linear Regression: The purpose of a linear regression analysis is to determine if one or more predictor variables can account for the presence or absence of a certain dependent (criterion) variable (StatisticsSolutions2013) calculated by (Hellings and Haelermans, 2022), (Smith et al., 2019), and (Cabada et al., 2020).
- Correlation: Correlation is a statistical term that reflects how much two or more variables change in relation to each other (Wigmore) which was calculated by (Malliarakis et al., 2017) and (Smith et al., 2019).

This systematic literature review provides valuable insights into the trends, best practices, and impacts of personalized gamified programming education on students' cognition. Most of the studies used various tools to evaluate programming language learning in their respective areas using questionnaires, interviews and observation methods. The findings of this research questions showed that most of the studies used questionnaire surveys and SPSS tools for the data analysis.

Table 12: Related studies evaluation measures

| Item No. | Ref. | Game | Measured Approach | Result Presented | Software |
|---|---|---|---|---|---|
| 1 | (Giannakoulas and Xinogalos, 2018) | Run Marco game | Effectiveness and Acceptance | Descriptive | SPSS |
| 2 | (Cabada et al., 2020) | EasyLogic | - | Descriptive and t-test, regression analysis | SEM |
| 3 | (Malliarakis et al., 2017) | CMX environment | Effectiveness | Descriptive, Mean, S. D Correlation | SPSS |
| 4 | (Papadakis, 2020) | Dr. Scratch | Evaluation | Mean, S. D | SPSS |
| 5 | (Topalli and Cagiltay, 2018) | Dr. Scratch | Improvement | t-test | SPSS |
| 6 | (Chang et al., 2020) | Programmer Adventure | Effectiveness | t-test | SPSS |

Continued on next page

Table 12: Related studies evaluation measures (Continued)

| | | | | | |
|---|---|---|---|---|---|
| | | Land | | | |
| 7 | (Jakoš and Verber, 2017) | Aladdin and his flying carpet | Improvement | Mean, S.D, Paired-samples t-test, ANOVA | SPSS, Excel |
| 8 | (Garneli and Chorianopoulos, 2018) | Dr. Scratch | Exploring | Non-parametric Wilcoxon signed-rank test (z) and non-parametric Mann–Whitney U test | SPSS |
| 9 | (Mathew et al., 2019) | PROSOLVE | Problem-Solving | Descriptive | SPSS |
| 10 | (Pellas and Vosinakis, 2018) | Scratch and OpenSim with the Scratch4SL palette | Effectiveness | Mean, S.D, Mann-Whitney U | SPSS |
| 11 | (?) | Computational Thinking with Scratch | Effectiveness | ANCOVA | SPSS |
| 12 | (Strawhacker and Bers, 2019) | Dr. Scratch | Investigation | Non-parametric Kruskal–Wallis H, or Kruskal–Wallis | SPSS |
| 13 | (Hitchens and Tulloch, 2018) | A software | - | - | - |
| 14 | (Syaifudin et al., 2020) | Android Programming Learning Assistance System | Test-driven development method | - | - |
| 15 | (Marwan et al., 2019) | iSnap | Evaluation | Interview | - |
| 16 | (Maskeliu¯nas et al., 2020) | - | Effectiveness | - | - |
| 17 | (Duffany, 2017) | - | - | - | - |
| 18 | (Seraj et al., 2018) | BEESM | - | - | - |
| 19 | (Figueiredo and García-Peñalvo, 2018) | - | - | - | - |
| 20 | (Krugel and Hubwieser, 2017) | MOOC called LOOP | Computational Thinking | Textual feedback | - |
| 21 | (Skalka and Drlík, 2018) | - | - | - | - |
| 22 | (Nadolny et al., 2017) | - | - | - | - |
| 23 | (Hooshyar et al., 2018) | Online Game-based Bayesian Intelligent Tutoring System | Evaluation | Interview | - |

Continued on next page

Table 12: Related studies evaluation measures (Continued)

| | | | | | |
|---|---|---|---|---|---|
| 24 | (Hausswolff, 2017) | - | - | - | - |
| 25 | (Drosos et al., 2017) | HappyFace | Identification | - | - |
| 26 | (Bernik et al., 2017) | - | - | - | - |
| 27 | (Troiano et al., 2019) | Dr. Scratch | Evaluation | Descriptive, Cluster analysis, and Data visualization | - |
| 28 | (Malik et al., 2019) | PROBSOL | Problem-Solving | - | - |
| 29 | (Devine et al., 2019) | MS MakeCode and CODAL | Evaluation | - | - |
| 30 | (Yallihep and Kutlu, 2020) | Lightbot | Effectiveness | Descriptive, t-test | SPSS |
| 31 | (Piedade et al., 2020) | - | Computational Thinking | - | - |
| 32 | (Luik et al., 2019) | - | - | - | - |
| 33 | (Luxton-Reilly et al., 2019) | - | Check Pass Rate | - | - |
| 34 | (Marín et al., 2018) | - | Problem-based learning | Descriptive | SPSS |
| 35 | (Smith et al., 2019) | - | Effect | Descriptive, Linear Regression, Correlation | SPSS |
| 36 | (Schez-Sobrino et al., 2020) | RoboTIC | Motivation | Descriptive | SPSS |
| 37 | (Ivanović et al., 2017) | LMS | Effectiveness | Kruskal-Wallis ANOVA, Mann-Whitney U Kolmogorov-Smirnov test | SPSS |
| 38 | (Hellings and Haelermans, 2022) | - | Effect | Descriptive, Regression | SPSS |
| 39 | (Marwan et al., 2019) | iSnap | Impact | Descriptive | SPSS |
| 40 | (Laporte and Zaman, 2018) | - | - | - | - |
| 41 | (Kumar and Sharma, 2019) | ProLounge | Achievement | Descriptive Results | Excel |
| 42 | (de Pontes et al., 2019) | - | - | - | - |

Continued on next page

Table 12: Related studies evaluation measures (Continued)

| | | | | | |
|---|---|---|---|---|---|
| 43 | (Paiva et al., 2020) | - | Impact | Descriptive, Mann-Whitney U one-sided tests | SPSS |
| 44 | (Wong and Yatim, 2018) | Odyssey of Phoenix | Learning | Descriptive, Paired Sample T-Test, ANOVA | SPSS |
| 45 | (Gulec et al., 2019) | CENGO | Achievement | Descriptive | - |
| 46 | (Tasadduq et al., 2021) | CYourWay | Effect | Descriptive, Independent Sample t-test, Mann-Whitney u test | SPSS |
| 47 | (Abbasi et al., 2021) | POOsg | Performance, Motivation | Descriptive, Paired t-test | SPSS |
| 48 | (Zhu et al., 2019) | Parallel | Effectiveness | Descriptive | SPSS |
| 49 | (Sideris and Xinogalos, 2019) | PY-RATE ADVENTURES | Learning | Descriptive | SPSS |
| 50 | (Montes et al., 2021) | DFD-C | Effectiveness | Descriptive | SPSS |
| 51 | (Xinogalos and Tryfou, 2021) | Game of Code: Lost in Javaland | Motivation | Descriptive | SPSS |
| 52 | (Toukiloglou and Xinogalos, 2022) | Dungeon Class | Effectiveness | Frequency, Kruskal–Wallis test | SPSS |
| 53 | (Daungcharone et al., 2017) | CPGame | Effectiveness | Mean, SD, MANOVA | SPSS |
| 54 | (Carreño-León et al., 2019) | - | Effectiveness | Frequencies | SPSS |
| 55 | (Jemmali et al., 2019) | May's Journey 3D puzzle game | Learning | - | - |
| 56 | (Khaleel et al., 2019) | Gami-PL | Effectiveness, Motivation | Mean, SD, t-test | SPSS |
| 57 | (Moreno and Pineda, 2018) | Gamification activities | Learning | Mean, SD, Skewness, Kurtis | SPSS |
| 58 | (Pankiewicz, 2020) | - | Impact | U Mann-Whitney test | SPSS |
| 59 | (Queirós, 2019) | PROud Framework | - | - | - |
| 60 | (Marín et al., 2018) | UDPiler | Investigation | Descriptive | SPSS |

## 6. Discussion and future directions

The consensus on the current state of the plethora of gamification in education research is that gamification consistently improves motivation and performance, as shown in Figure 3. The results of this systematic literature review have shed light on the importance of gamification, personalization, and cognition in programming language education. The findings suggest gamification techniques enhance programming education engagement, motivation, and learning outcomes. Personalization of gamified programming education has also been identified as a key factor in improving student performance and satisfaction. Moreover, the results have shown that gamification can be tailored to different cognitive levels of Bloom's taxonomy to promote higher-order thinking skills. Personalized gamification frameworks can also help students learn at their own pace and provide a more enjoyable and rewarding learning experience. Furthermore, programming language education can be enriched using various gamification techniques, such as game elements, game design principles, and game-based learning approaches. The results also suggest that different programming languages require different gamification strategies to be effective.

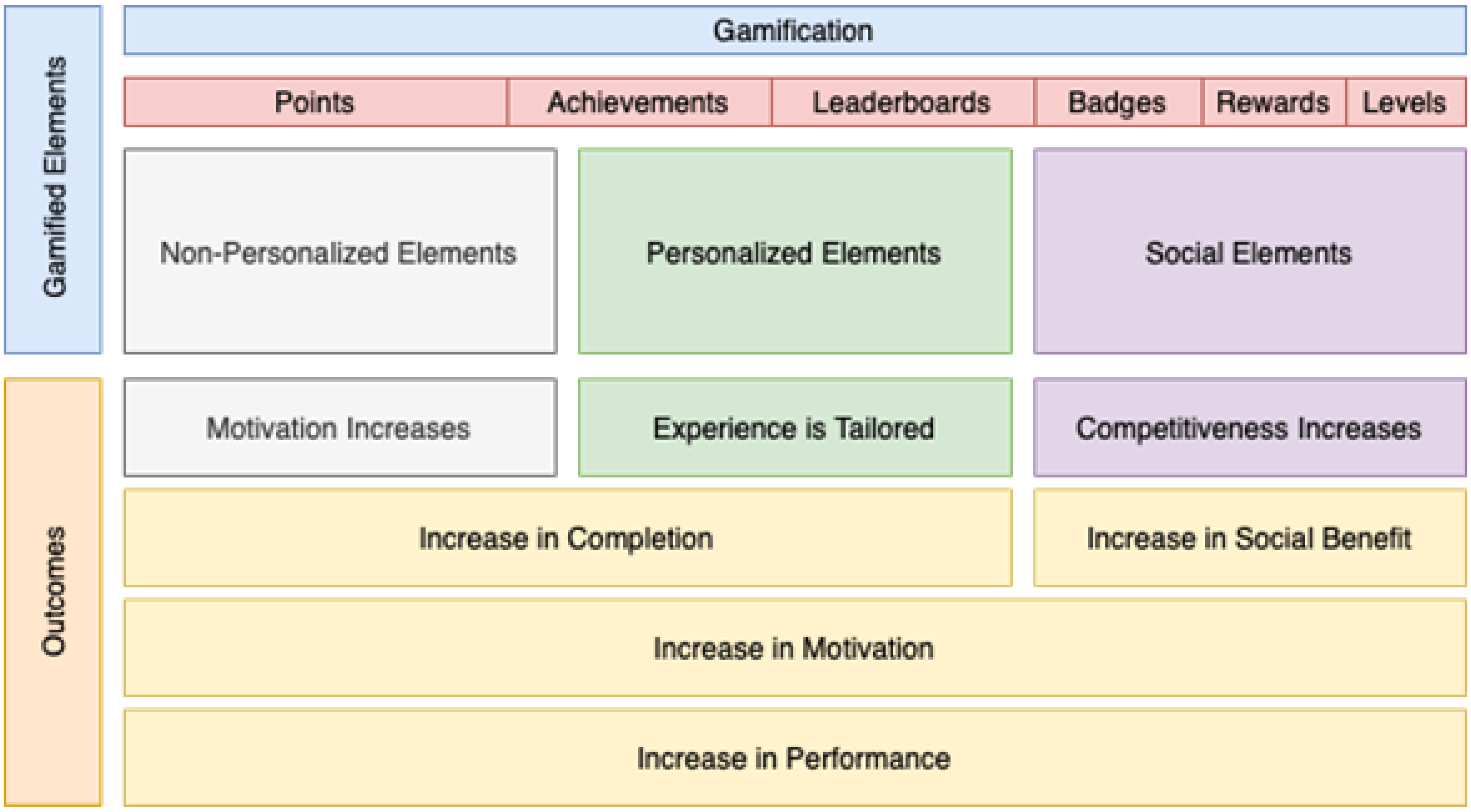

**Figure 3:** Diagram of gamification elements and their outcomes.

In conclusion, the findings of this systematic literature review indicate that gamification and personalization are promising strategies for enhancing programming language education. The results also highlight the importance of considering cognitive factors when designing gamified programming education. Further research is needed to explore the effectiveness of different gamification strategies in various programming languages and to evaluate the impact of personalized gamified programming education on student learning outcomes.

## 7. Findings, Challenges, and Recommendations

The systematic literature review revealed several key findings regarding personalized gamified programming education. First, it was found that personalized gamification strategies can improve student engagement and motivation in programming education. Second, personalized gamification can enhance students' problem-solving and cognitive abilities. Third, gamified applications' design and customization can significantly impact personalized gamification strategies' effectiveness. Fourth, there is a need for more empirical studies to validate the effectiveness of personalized gamification strategies in programming education. Finally, the review identified a lack of consensus on the evaluation criteria and metrics for assessing the quality of personalized gamification applications in programming education.

Several challenges were identified during the systematic literature review. One of the primary challenges is the limited availability of high-quality research on personalized gamified programming education. Additionally, the lack of standardization in designing and evaluating gamified applications makes comparing the effectiveness of different personalized gamification strategies difficult. Another challenge is the need for skilled instructors who can effectively design and implement personalized gamification strategies in programming education.

Based on the findings and challenges identified in this systematic literature review, the following recommendations are made:

- More empirical studies are needed to validate the effectiveness of personalized gamification strategies in programming education.

- Standardizing the design and evaluation of gamified applications can enhance the comparability of different personalized gamification strategies.
- More emphasis should be placed on developing tools and resources to support instructors in designing and implementing personalized gamification strategies in programming education.
- Future research should focus on identifying the most effective personalized gamification strategies for different cognitive levels and programming languages.
- There is a need for consensus on the evaluation criteria and metrics for assessing the quality of personalized gamification applications in programming education.

## 8. Conclusion

In conclusion, this systematic literature review highlights the significance of gamification, personalization, cognition, and programming education in enhancing students' learning outcomes. The review identifies several trends and best practices for implementing personalized gamification frameworks in programming education and highlights the benefits of gamified programming education in promoting students' cognitive abilities.

However, this review also reveals some challenges associated with gamification and personalization in programming education, such as the need for appropriate tools and software applications tailored to specific programming languages and concepts. Additionally, the review identifies some gaps in the literature, such as the limited research on the long-term effects of personalized gamified programming education and the lack of studies evaluating the effectiveness of gamification across different programming languages.

To overcome the challenges and gaps in programming education, we suggest that researchers and educators concentrate on creating tailored gamification strategies and software tools that cater to the specific requirements of programming learners. It is also crucial to conduct more research to examine the long-term effects of gamified programming education and assess the effectiveness of gamification across various programming languages. Furthermore, game-based learning has demonstrated its effectiveness in several domains beyond programming education, such as language learning, where games can offer a fun and interactive way for learners to practice their language skills. Other domains like healthcare, business, and marketing can benefit from game-based learning by providing learners with realistic simulations and experiential learning. Lastly, it is vital to explore the function of cognitive factors in gamified programming education and incorporate them into developing tailored gamification frameworks.